\documentclass[prb,twocolumn,superscriptaddress]{revtex4}
\usepackage{graphicx}
\usepackage{dcolumn}
\usepackage{bm}
\usepackage[latin1]{inputenc}
\usepackage{amsfonts}
\usepackage{amssymb}
\usepackage{amsmath}
\begin{document}
\title{First-principles calculations of the magnetic properties of (Cd,Mn)Te nanocrystals}      

\author{C. Echeverr\'ia-Arrondo}
\affiliation{Departamento de F\'isica, Universidad P\'ublica de Navarra, E-31006, Pamplona, Spain}
\affiliation{Donostia International Physics Center (DIPC),  E-20018, San Sebasti\'an/Donostia, Spain}
\affiliation{Departamento de F\'isica de Materiales, Facultad de Qu\'imicas, Centro de F\'isica de Materiales CSIC-UPV/EHU, E-20018, San Sebasti\'an/Donostia, Spain}

\author{J. P\'erez-Conde}
\affiliation{Departamento de F\'isica, Universidad P\'ublica de Navarra, E-31006, Pamplona, Spain}

\author{A. Ayuela}
\affiliation{Departamento de F\'isica de Materiales, Facultad de Qu\'imicas, Centro de F\'isica de Materiales CSIC-UPV/EHU, E-20018, San Sebasti\'an/Donostia, Spain}
\affiliation{Donostia International Physics Center (DIPC),  E-20018, San Sebasti\'an/Donostia, Spain}

\date{\today}
\begin{abstract}
 We  investigate the  electronic and  magnetic properties  of Mn-doped
CdTe  nanocrystals (NCs)  with $\sim2$  nm  in diameter  which can  be
experimentally   synthesized   with  Mn   atoms   inside.   Using   the
density functional   theory,  we  consider   two  doping   cases:  NCs
containing one or two Mn impurities.  Although the Mn \textit{d} peaks
carry five up  electrons in the dot, the local  magnetic moment on the
Mn site is 4.65 $\mu_{B}$. It is smaller than 5 $\mu_B$ because of the
\textit{sp-d}   hybridization   between   the  localized   3\textit{d}
electrons  of the  Mn atoms  and the  \textit{s}-  and \textit{p}-type
valence states of the  host compound.  The \textit{sp-d} hybridization
induces small  magnetic moments on  the Mn nearest-neighbor  Te sites,
antiparallel to Mn moment affecting the \textit{p}-type valence states
of the  undoped dot, as usual for a kinetic mediated exchange magnetic coupling.  Furthermore,  we calculate the parameters standing
for  the \textit{sp-d}  exchange interactions:  Conduction $N_0\alpha$
and valence $N_0\beta$ are close  to the experimental bulk values when
the Mn  impurities occupy bulk-like  NCs' central positions,  and they
tend to zero close to the surface.  This behavior is further explained
by  an analysis  of  valence band-edge  states  showing that  symmetry
breaking splits  the states and  in consequence reduces  the exchange.
For two Mn atoms in several  positions, the valence edge states show a
further  departure  from an  interpretation  based  in a  perturbative
treatment.  We  also calculate the  \textit{d-d} exchange interactions
$\vert J^{dd}\vert$ between Mn spins.   The largest  $\vert J^{dd}\vert$ value  is also
for  Mn atoms  on  bulk-like  central sites;  in  comparison with  the
experimental     \textit{d-d}     exchange     constant    in     bulk
$\rm{Cd}_{0.95}\rm{Mn}_{0.05}Te$ it is four times smaller.
\end{abstract}
\maketitle

\section{Introduction}


The recent progress  in chemical synthesis, computational capabilities
and scanning-probe  techniques has permitted  a detailed understanding
of semiconductor  nanocrystals (NCs)  -- also known  as nanoparticles,
clusters, crystallites or quantum  dots (QDs).  Their properties which
depend  on  size  \cite{wang42,wang,albe58,perez110,sapra}  and  shape
\cite{albe58}  have overimposed  effects due  to  quantum confinement.
For instance,  when the QD radius  is smaller than the  Bohr radius of
exciton, quantum-confinement effects appear, such as the blue shift of
gaps        and       the        discretization        of       energy
spectra\cite{wang42,wang,albe58,perez110,sapra,albe57}.     The    gap
properties     can     also     be    tuned     intentionally     with
doping\cite{albe57,bhatta83,bhargava,bhatta0,bhatta1,norris,melnikov,erwin}.
Recently,  much effort has  focused on  II-VI semiconductor  NCs doped
with  magnetic impurities  such  as Mn,  which  is the  topic of  this
work. The  Mn doping is  motivated by diluted  magnetic semiconductors
(DMSs).   These   compounds  are   bulk   semiconductors  doped   with
transition-metal  impurities  such  as  Cr,  Mn,  Fe  and  Co  at  low
concentrations.  DMSs  are current research  materials in spintronics,
integrated    in    novel    magnetoelectronic   devices    such    as
spin-LEDs\cite{wolf}.   Their remarkable magnetic  and magneto-optical
properties result from  the strong \textit{sp-d} exchange interactions
between band  carriers and Mn  ions\cite{bhatta1}.  These interactions
yield  giant  band-edge  splittings  at  low  temperature  (about  100
meV)\cite{szczytko}.   In particular,  little  is known  theoretically
about exchange in NCs made of II-VI DMSs doped with Mn.

DMS NCs of type II-VI doped with Mn have been successfully synthesized
and characterized during the last  fifteen years. These works found at
zero  field,  or low  fields,  contradictory  results  for the  Zeeman
splitting.   In comparison with  bulks, several  authors have  shown a
reduction   of   the    excitonic   Zeeman   splitting   in   Mn-doped
CdS\cite{hoffman},                CdSe\cite{yanata}                and
CdTe\cite{bhatta1,bhatta2,merkulov}  NCs.   Other  works  revealed  no
enhancement of  the excitonic Zeeman  splitting in Co-doped  ZnO, ZnSe
and  CdS  QDs\cite{gamelin}  in  comparison with  their  bulk  values.
Magnetic circular dichroism and optical experiments with Mn-doped ZnSe
NCs revealed  a value for  the same splitting  of 28 meV,  much larger
than in bulk ZnSe:Mn \cite{norris}.
  
The Zeeman splittings are due  on one hand to the confinement-enhanced
\textit{sp-d}  hybridization  between   the  occupied  Mn  3\textit{d}
orbitals and the \textit{sp}-type valence states of the host compound;
and  on  the  other  hand  to  the crystal  field  experienced  by  Mn
impurities,  which is  significantly different  nearer the  NC surface
than  in bulk\cite{kennedy}.   Experimentally,  several authors  found
that  Mn atoms  are embedded  in ZnSe  NCs \cite{norris},  and  even a
single  Mn  impurity  is  inside  CdTe  nanocrystals  \cite{besombes}.
Theoretically, enhanced splittings are  obtained in ZnSe:Mn NCs within
the effective  mass approximation\cite{bhatta3} with a single Mn sitting
at the center.  However, there is little theoretical information about
exchange interactions  and splittings for CdTe:Mn NCs, and  even less
with Mn  off-center.  We  also note that  the DMS NCs can  hold inside
several Mn impurities.

In the present work we study Mn-doped CdTe NCs of spherical shape with
the density functional theory.  In section \ref{computational} we give
a brief account of the  theorical framework and numerical details. The
host compound  is a well-known  wide-gap semiconductor of  type II-VI,
and manganese is a widely-used dopant, known for activating photo- and
electroluminescence,   and   also   for  contributing   to   efficient
luminescence  centers (3\textit{d}  electrons)  \cite{huong}.  Several
authors have  previously calculated the properties of  bulk II-VI CdTe
doped  with  Mn\cite{larson, deportes, merad}.  We consider  two  doping
situations: NCs including one  and two Mn impurities, which substitute
for one or two Cd atoms  in the zinc blende lattice, respectively.  In
section \ref{nanosingle}  the results concerning NCs with  a single Mn
impurity  are given.   The  results  for NC  with  two impurities  are
described   in   section  \ref{nanotwo}.   We   calculate  the   total
ground-state  energy of the  QD, for  several positions  of impurities
within the  crystal and the  magnetic state.  Moreover, we  obtain the
\textit{sp-d}  and \textit{d-d}  exchange constants\cite{larson}  as a
function of the Mn locations: $N_0\alpha$ and $N_0\beta$ stand for the
exchange interactions between the Mn local moments and the \textit{s}-
and  \textit{p}-type  band-edge  states  of the  host  CdTe;  $J^{dd}$
parametrizes the exchange interaction  between two Mn spins.  Finally,
we  sum  up  the  main  findings and  concluding  remarks  in  section
\ref{conclusion}.   The  system  described  here may  provide  further
understanding of  solid-state qubits, since  it permits to  detect and
manipulate a  single spin\cite{besombes}.   In addition it  could show
magnetic  and magneto-optical properties,  such as  fast recombination
and high luminescence efficiency \cite{bhargava}.


\section{Computational Details}
\label{computational}
The many-body  problem for the  electrons around the nuclei  is solved
based  on the  density functional  theory (DFT),  using  the Kohn-Sham
equations.   The  valence electrons  move  in  the external  potential
created by the nuclei and the core electrons. The electronic states of
the studied NCs are  obtained from the projector augmented-wave method
as  implemented in the  VASP code  \cite{kresse1,kresse2,kresse3}.  To
account  for  the  \textit{sp-d}  hybridization\cite{larson}  the  ten
3\textit{d}  spin-orbitals  of  Mn  impurities  are  included  in  our
calculations.  Within  DFT, our approach  for the exchange-correlation
is  defined  with  the  generalized-gradient  approximation  (GGA)  of
Perdew, Burke  and Ernzerhof\cite{perdew}.  However,  the interactions
among  the  $3\it{d}   $  electrons  of  Mn  impurities   in  the  GGA
approximation are  only partially  described, since they  are strongly
localized.  Thus,  we use  the  so-called  GGA+\textit{U} scheme  and
introduce  in  the  calculations  two  common  correction  parameters,
\textit{U} and  \textit{J} \cite{freeman}. For Mn atoms  in CdTe bulk,
$U=6.2$  eV and  $J=0.86$  eV.   They are  considered  to correct  the
Coulomb (\textit{U}) and  exchange (\textit{J}) interactions among the
3\textit{d} electrons at Mn sites\cite{freeman,solovyev}. We have checked that with the $U$ values of Refs.~\onlinecite{solovyev} and \onlinecite{freeman} for Mn, the density of states are very similar, especially the $d$ peak corresponding to Mn is nearly at the same energy. This peak is below the $p$-type part of the CdTe valence band. We have chosen the recent $U$ value in Ref.~\onlinecite{freeman} because it calculated MnTe bulk where Mn is surrounded by Te. Additionally, tests on bulk MnTe with the same lattice constant of Ref.~\onlinecite{freeman} show nearly the same band structures when using LSDA or GGA and the same $U$ value.

The input  parameters for  the VASP calculations  are determined  in a
preliminary  work for  two model  bulk  systems, CdTe  and F-MnTe (zinc blende).   For
several lattice constants  we fitted the ground-state  energies to the
Murnaghan's  equation  of  state,  which  depends  on  the  unit  cell
volume. The equilibrium lattice constants of bulks CdTe and F-MnTe
  are  $a_{CdTe}=6.63$  ~{\AA}   and  $a_{F-MnTe}=6.38$
~{\AA}. The cut-off energy in the plane wave basis set is 350 eV; this
is the  shared cut-off to  converge the ground-state energies  of both
CdTe and F-MnTe bulks  within meV.

The studied NCs are spherical  about $17$~{\AA} in diameter centered
on a cation site (Cd).  The crystal
structure  is  zinc blende,  which  has  tetrahedral ($T_d$)  symmetry
around the atoms. These QDs are passivated since otherwise the surface
dangling  bonds  would  introduce   surface  states  in  the  near-gap
spectrum. Organic ligands  are commonly used to passivate  NCs, but we
want    to   center    on    the   intrinsic    properties   of    NCs
\cite{chelikowsky}.  For  the sake  of  simplicity  we  resort to  the
simplest  passivation agent:  a pseudohydrogen  atom  (H$^\ast$).  The
fictitious H$^\ast$  is characterized by  fractional electronic charge
and by a  proton with the same charge but  positive. Every Cd (5$s^2$)
dangling bond  includes 2/4  unpaired electrons, so  it is bound  to a
pseudohydrogen    atom    with   a    fictitious    charge   of    6/4
\textit{e}. Similarly, every Te  dangling bond ($5s^2\;5p^4$) is bound
to a pseudohydrogen atom with an electronic charge of 2/4 \textit{e}. When the pseudohydrogen atoms H$^\ast$
are included the NCs  have  107 atoms in total: $19$ Cd, $28$ Te and $60$ H$^\ast$.
 We dope them with
one or two  Mn impurities depending on the  studied case.

\begin{figure}[t]
\includegraphics[scale=0.7]{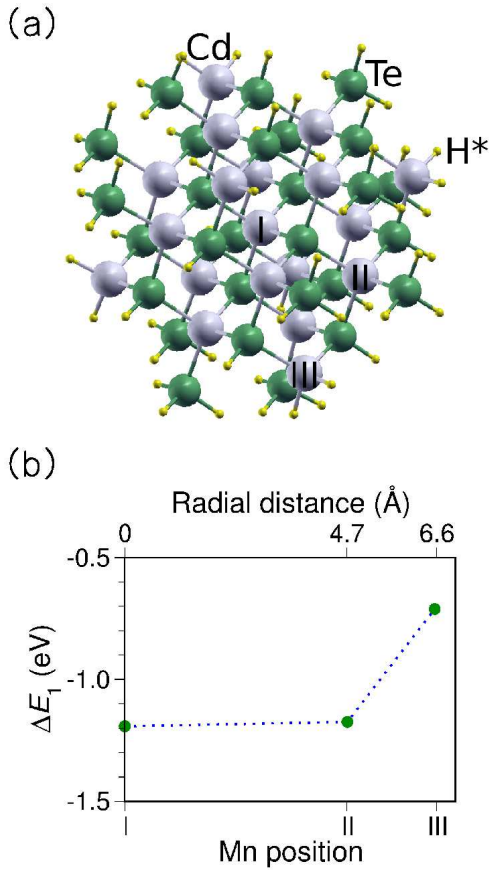}
\caption{\label{fig:esfera}(Color   online)    (a)   Nanoparticle   of
(Cd,Mn)Te with 107 atoms: Cd atoms with light gray, Te atoms with dark
gray  (green)  and   passivating  $\rm{H}^{\star}$  atoms  with  small
spheres. The cation sites are distributed in three groups; labelled as
"I"  for  the center,  "II"  for inner  atoms  and  "III" for  nearest
neighbors to surface atoms. (b) Substitutional energy $\Delta E _1$ for a single Mn impurity.
We are referring to the ions in solution. The higher is the energy the more stable is the Mn atom in such position.  }
\end{figure}

We   use  the   supercell  approximation   to  deal   with   a  single
nanocrystal. We place the NC at  the center of a large cubic unit cell
and take the  $\Gamma$ point. Then we have  converged the ground-state
energies of two small model NCs ( $\rm{Cd}\rm{Te}_{4}\rm{H^\ast}_{12}$
and  $\rm{Mn}\rm{Te}_{4}\rm{H^\ast}_{12}$) versus  the  supercell size
within meVs.  We consider that  this separation between walls is valid
for larger  NCs.  The input  distances between nearest-neighbor atoms
are    taken     from    those     in    Cd-Te    bulk.     We    have
$d_{Cd-Te}=(\sqrt{3}/4)a_{CdTe}=2.87$~{\AA}          and          take
$d_{Mn-Te}=d_{Cd-Te}$.   The atomic structures  are relaxed  until the
forces on each atom are smaller than 0.02 eV/~{\AA}.


\section{Nanocrystals with a single Mn impurity}
\label{nanosingle}


\subsection{Energetics and Geometry for Mn Atom in Different Positions}

We study NCs doped with a  unique Mn impurity substituting a single Cd
atom. The  NC geometry  is plotted in  Fig. \ref{fig:esfera}  (a). The
cation sites of  the NC are distributed in three  sets, and labeled as
"I" for  the sphere center, "II"  for other inner  positions and "III"
for  outside  positions.   The  substitutional energy  is  the  energy
difference $\Delta E_1$ of the following reaction:
\begin{equation}
\begin{array}{ccc}
\textrm{CdTe}\;\textrm{NC}+\textrm{Mn}^{+2} & \longrightarrow & \textrm{CdTe:Mn}\;\textrm{NC}+\textrm{Cd}^{+2}.
\end{array}
\end{equation}
Due to the  preparation of NCs in solution, we  take as source systems
for Mn and  Cd their respective ions. The  substitutional energies are
shown in Fig.  \ref{fig:esfera} (b) for the positions  I, II, and III.
All the  energies are  endothermic. This means  that the  synthesis of
these  NCs   requires  high  temperatures  as   shown  in  experiments
\cite{norris}. The changes in $\Delta E_1$ show opposite trends to the QD
calculated total energies as a  function of Mn site. The lowest-energy
sites belong  to group  III with  Mn atoms close  to the  surface; the
outward sites  are the  most stable positions  for the  impurity. This
outward  stability  is  consistent  with other  results  for  magnetic
impurities in other semiconductor nanoscale objects, such as nanowires
\cite{nanowire}.

In  the output  geometry, the  nearest-neighbor distances  to the  Mn
impurity  are about  2  \% smaller  than  the Cd-Te  distance for  the
undoped  NC.   This  contraction   effect  depends  strongly   on  the
exchange-correlation approach. The neighbor Mn-Cd distances at the GGA
level contract by  more than 4 \% respect to  the bulk Cd-Te distance. However,
when relaxing the  geometries with the +$U$ scheme  the distances expand
closer to  the unrelaxed input Cd-Te distances.  This result could justify
the use of  unrelaxed Cd-Te distances in other  works about Mn doping,
but in principle this agreement seems fortuitous.

\subsection{ Magnetism and Electronic Properties}

\begin{figure}[t]
\centering
\includegraphics[scale=0.9]{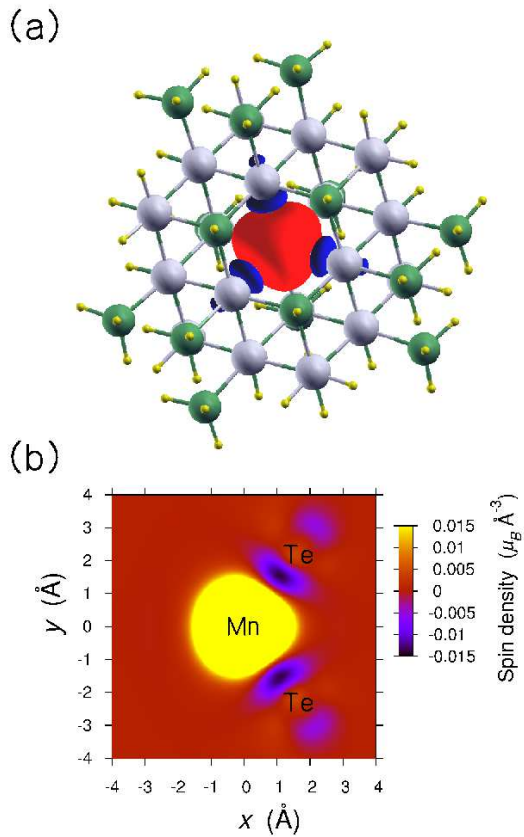}
\caption{\label{fig:spin-density}(Color  online) (a) Spin-polarization
isosurfaces when the Mn impurity is placed at the center for cuts at $\pm0.01$
$\mu_{B}/$~{\AA}$^3$. Positive  values are light  gray (red); negative
values,  dark  gray (blue).   (b)  Spin  density  in a  plane 
defined  by   the central Mn   impurity and  two
nearest-neighbor Te atoms. For the sake of clarity, spin-up density is
chopped at 0.015  $\mu_{B}$~{\AA}$^{-3}$. The spin-down density region
yields  on  the  Te  sites  a  magnetic  moment  of  -0.01  $\mu_{B}$,
antiferromagnetically  coupled   to  the  Mn   magnetic  moment  (4.65
$\mu_{B}$).}
\end{figure}

 We analyze next the origin of magnetism as we are dealing with Mn, that typically is a magnetic element. The  total  magnetic  moment associated with the QD is 5 $\mu_{B}$, as expected since the Mn dopant
introduces  five spin-up electrons.  The local  magnetic moment  in Mn
impurity   is  nevertheless  smaller,   $4.65~\mu_{B}$.   It  differs
substantially from the  bulk  value, $4.21~\mu_{B}$,   for  Mn in bulk  CdTe as  given in a  recent work
\cite{merad}.    The   spin   density   is   spatially   plotted   in
Fig. \ref{fig:spin-density}. The negative  areas of
spin density  in the Mn  nearest-neighbor Te sites integrate  to small
magnetic  moments   about  $-0.01$  $\mu_{B}$.  Such   Te  atoms  were
non-magnetic before doping. Other  radii for the  integration spheres  different from  the default
Wigner-Seitz radii  change only  slightly these local  magnetic moment
values. Note that the +$U$  approach affects the charge distribution and
thus  modify  the local  moment  on Te,  which  changes  in sign  when
improving  the  $d$-level  description.    This  behavior  is  in disagreement with other previous results concerning CdTe:Mn bulk \cite{merad}, where the Te and  Mn moments are found to be  parallel.  The Mn  magnetic moment is lower
than 5  $\mu_{B}$ because  of the \textit{sp-d}  hybridization between
the localized 3\textit{d} electrons of Mn impurity and the delocalized
\textit{s}- and \textit{p}-type valence states of CdTe host.

Before studying the $sp$-$d$  hybridization, lets comment the gaps using
the  local density  of states  (DOS) as  given in  Fig. \ref{fig:dos}.
Quantum confinement  in QDs produces a  blue shift of the  gap and all
the related optical properties,  such as excitons. The gaps calculated
using LDA  or GGA approximations  are well known to  underestimate the
experimental values.  The calculated gap of CdTe bulk is 0.69 eV which
is lower than in experiments, 1.48 eV. Our calculated HOMO-LUMO gap in
the Cd-Te  QD is 2.59  eV, which shows  the predicted blue  shift. All
these findings  are well  established in semiconductor  QDs.  However,
the HOMO-LUMO  gap of  the CdTe dot  doped with  Mn and within  the +$U$
approach remains  almost constant to  2.62 eV [see  Fig. \ref{fig:dos}
(a)]  when  Mn  is in  the  center.  We  note  that the  GGA  approach
understimates  the gap and becomes  2.04  eV. Moving  Mn to  other
positions  changes this  gap  value within  a  tenth of  eV, which  is
negligible to be commented [see Fig. \ref{fig:dos} (b)].

\subsubsection{Origin of Mn QD-Exchange Coupling}

\begin{figure}[t]
\centering
\includegraphics[scale=0.7]{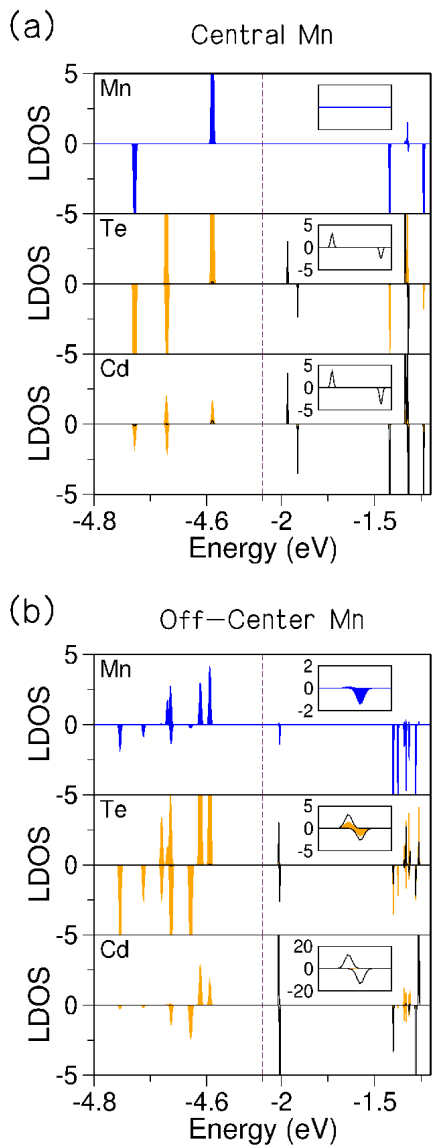}
\caption{\label{fig:dos}(Color online) Projected DOS in Mn impurity, $d$-type, and in the  nearst-neighbor atoms Cd  and Te, $s$-type in black (black) and $p$-type in gray (orange). In panel (a)  Mn is
placed at the QD center; and panel  (b), Mn at site II. Insets widen the conduction edge states. The dashed vertical lines separate the edges of valence band and conduction band. }
\end{figure}

The \textit{sp-d} hybridization is  studied using the local density of
states  (LDOS) projected  into  the  orbitals of  Mn  impurity and  of
nearest-neighbor Te atoms.  The main peak of Mn $d$-states appears below
the  $p$ part valence  band.  Due  to the  tetrahedral symmetry  in CdTe
lattice, the  five spin-up  Mn electrons are  split into a  triplet of
$t_2$  symmetry  and  a   doublet  of  \textit{e}  symmetry.   We  are
interested in  the magnetic coupling  between Mn and  NC
states equivalent  to the band-edge  states. The projected  DOS around
valence- and  conduction-edge states  are shown in  Fig. \ref{fig:dos}.
The   conduction-edge   states   of  the   host  compound   show
\textit{s}-type character  and do  not hybridize with  the 3\textit{d}
orbitals  of   central  Mn,  or  hybridize   sligthly  for  off-center
Mn. Hence,  the \textit{s-d}  exchange interaction arises  mainly from
Coulomb     repulsion      and     the     Pauli     exclusion
principle\cite{bhatta83,bhatta0,szczytko}, and originates the spin splitting of
these   states.   This   splitting   is  thus   always
ferromagnetic, and  its exchange constant $N_0\alpha$  is positive. On
the  contrary,   the  valence  band-edge   states  of  the   CdTe  are
\textit{p}-type and allowed to hybridize fully with the Mn 3\textit{d}
orbitals.  Anyhow such \textit{p-d}  hybridization is small and yields
an  effective  exchange  mechanism  of interaction  between  Mn  atoms
\cite{larson}. This  exchange is related to  the opposite polarization
between  the   valence  band-edge   states  and  the   Mn  3\textit{d}
orbitals. The  \textit{p-d} interaction originates  the spin splitting
of valence-band edge and it is always antiferromagnetic, since all the
3\textit{d}  spin-up  states are  occupied  and  only  jumps into  the
3\textit{d} spin-down states are available. The corresponding exchange
parameter $N_0\beta$ is thus always negative.
\begin{figure}[t]
\centering
\includegraphics[scale=0.3]{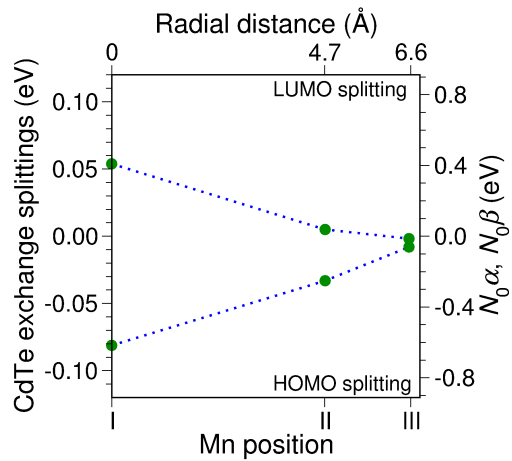}
\caption{\label{fig:dos2}(Color online) 
Exchange splitting of band edges (left) and the $N_0\alpha$ and $N_0\beta$ exchange constants (right) as a function of the position (below) and the distance (above) between the impurity and the NC center. Positive values are LUMO splittings; and negative values, HOMO splittings.}
\end{figure}

\subsubsection{ Site Dependence of the Exchange Constants }

Now  we  compute  the  \textit{sp-d} exchange  interaction  parameters
$N_0\alpha$ and  $N_0\beta$ following the expressions  for bulks. They
are     defined     in      the     standard     mean-field     theory
as\cite{bhatta83,larson,merad,gamelin,ayuela1,ayuela2}
\begin{equation}
N_0\alpha=\frac{\bigtriangleup E^c}{x\langle  S_z\rangle},\; \textrm{and} \; \;
N_0\beta=\frac{\bigtriangleup E^v}{x\langle S_z\rangle}.
\end{equation}
The  number $N_0$ is  the cations  per unit  volume\cite{bhatta3}. The
differences    $\bigtriangleup   E^{c,v}   =E^{c,v}$(spin    down)   -
$E^{c,v}$(spin  up)   are  the  spin  splittings   of  the  conduction
(\textit{c})-  and  valence  (\textit{v})-band  edges.   They  can  be
extracted   from   the   density    of   states   depicted   in   Fig.
\ref{fig:dos}.   The  \textit{x}  value   is  the   fractional  dopant
concentration  \cite{fatah}, and  $\langle  S_z\rangle=\frac{5}{2}$ is
the average \textit{z} component of Mn spins.

Following  eqn  (2)  we  calculate  $N_0\alpha$ and  $N_0\beta$  as  a
function   of    the   impurity   position.   They    are   shown   in
Fig.  \ref{fig:dos2}.  When  the  Mn  atom  occupies  the  NC  center,
$N_0\alpha=0.41$ eV,  $N_0\beta=-0.62$ eV and $N_0(\alpha-\beta)=1.03$
eV. The dopant concentration  is $x=1/19\sim0.05$ and these values are
comparable   with   the    experimental   ones   obtained   for   bulk
$\rm{Cd}_{0.95}\rm{Mn}_{0.05}Te$:         $N_0\alpha=0.22$         eV,
$N_0\beta=-0.88$ eV  and $N_0(\alpha-\beta)=1.10$ eV\cite{larson}. For
NCs, the  absolute values  of $N_0\beta$ and  $ N_0\alpha$  are close to each other.
 When  the impurity is located off-center, the  exchange constants are  smaller than  the
bulk values and they tend  to zero as the impurity  approaches the surface. This
finding  is explained  because  the Mn  bond  is expected  to be  less
covalent when  Mn is located near  the surface, where it  is less bulk-like. This  site dependence  of the exchange  constants seems  to have
implications  for  random  distribution   of  NCs.  When  there  is  a
collection of QDs,  and each containing a single  Mn impurity randomly
placed, the average $N_0(\alpha-\beta)$ would be significantly reduced
in comparison  with the bulk  intrinsic Zeeman splitting. This decrease  fits in
the   previous  theoretical   results   \cite{bhatta1,merkulov}.   The
decrease of the exchange splitting  is correlated with a smaller CdTe
dot  density  around Mn.  Thus,  a further  density analysis  of
band-edge states will be needed.

\subsubsection{Densities of HOMO-LUMO States, in and off-Center Positions}

\begin{figure}[t!]
\centering
\includegraphics[scale=0.9]{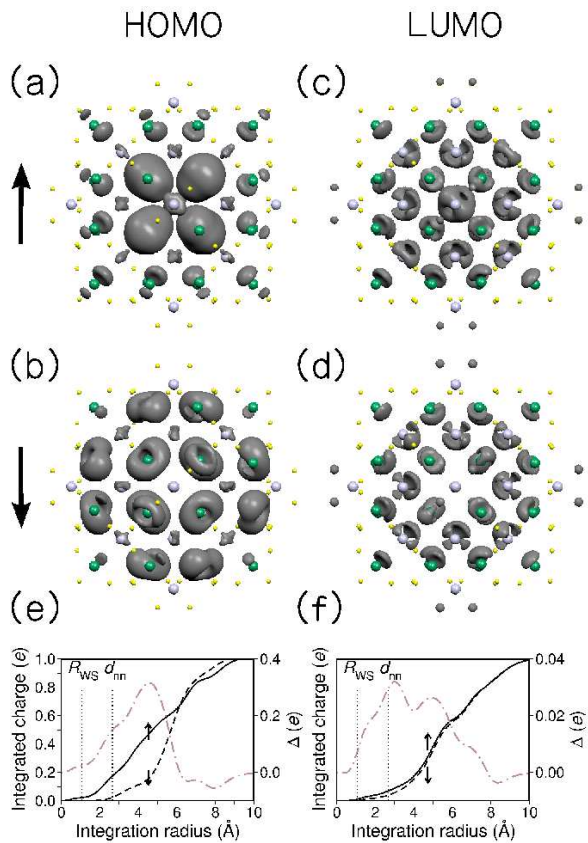}
\caption{\label{fig:HL}(Color   online)  Orbital densities  in the Mn-centered sphere for the  Highest
Occupied Molecular Orbital  (HOMO): (a) and (b) for up and down states. Also the densities for the Lowest Unoccupied Molecular Orbital are shown in (c) and (d) panels, up and down respectively. The  density cut is around
one  third  of  the  maximum  value.  The  integrated  charges  around
nanoparticle centers  are given  in (e) and  (f) for HOMOs  and LUMOs,
respectively.  Their up-down differences are plotted as dashed-dotted lines in gray. The Cd
Wigner-Seitz  Radius  $R_{\textrm{WS}}$   and  the  nearest-neighbor  distance
$d_{\textrm{nn}}$ are also shown with vertical dotted lines.}
\end{figure}

\begin{figure}[t]
\centering
\includegraphics[scale=0.85]{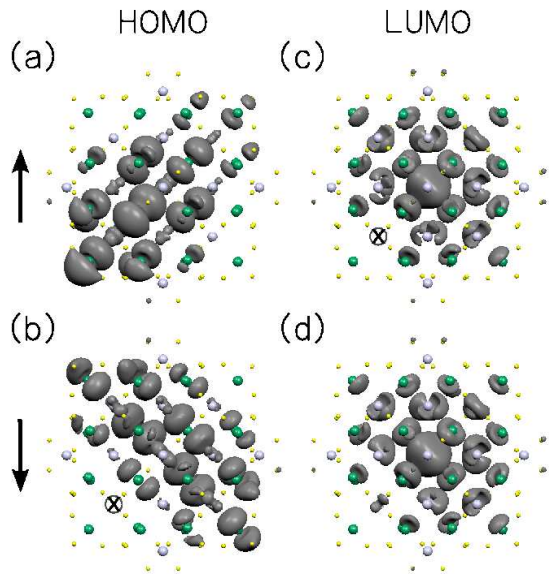}
\caption{\label{fig:HLII}(Color online) HOMO-LUMO densities for Mn off-center in position II. Panels (a)-(d) are organized as in the previous figure. The crossed atom denotes Mn.}
\end{figure}

Though the detailed  wavefunctions, as we have seen,  are not required
in  the study  of  exchange interaction  parameters,  we must  clearly
attribute them  to the  electron and the  hole effective  states which
reflect the character of the  conduction and valence states in the NCs
doped with  Mn. Our  idea is  to look at  the High  Occupied Molecular
Orbital (HOMO)  as a  representation for the  hole; and at  the Lowest
Unoccupied Molecular Orbitals (LUMO), for the electron. They are shown
in Figs. \ref{fig:HL} and \ref{fig:HLII}.

We do not include here the  HOMO-LUMO states of the undoped NC because
they  are  indistinguishable by  simple  eye  inspection  from the  up
wavefunction  in Fig.  \ref{fig:HL}  (a) and  (c)  for Mn  in the  dot
center.   The  main  contribution  to  the  up  HOMO  comes  from  the
nearest-neighbor  Te  atoms and  is  larger  than  from any  other  Te
atom. Although more delocalized,  the up LUMO has larger contributions
in the Cd atoms and in the center site, which is Mn or Cd for doped or
undoped dots respectively.  These spatial distributions reflect the Te
and Mn  local DOS  character for  the HOMO and  LUMO commented  in the
previous sections.

However, the HOMO and LUMO  down states are different from the undoped
NCs .   For the down states  we see that  the Mn placed in  the center
expels charge, which is quantified by integrating it around the sphere
center in Figs.  \ref{fig:HL} (e)  and (f). This effect is much larger
for the  HOMO than for  the LUMO, see  insets. This difference  can be
explained  partially because the  HOMO state  is occupied,  and mainly
because the LUMO state does not hybridize with the Mn states, as shown
in the previous DOS plot [Fig. \ref{fig:dos}], which means that it has
lower interaction  with the $d$-electrons  of Mn. The down  HOMO state
undergoes  the  strongest  change   in  comparison  with  its  undoped
counterpart   because   the   undoped    HOMO   is   mainly   in   the
nearest-neighbor Te to the central  Mn, with which the Mn must couple
antiferromagnetically.

To end this section we comment the density of the HOMO and LUMO states
for  off-center  Mn.   They  are   plotted  for  the  position  II  in
Fig. \ref{fig:HLII}. There  is larger mixing in the  LUMOs which makes
that both up  and down states remove charge  from the Mn neighborhood.
The  general form  of the  rest of  the LUMOs  resembles those  of the
sphere  without  Mn.  The  HOMOs  with  off-center Mn  suffer  larger
differences with respect  to the undoped case, specially  for the down
component.  Anyhow, we  see a  depletion of  charge around  central Mn
mainly for  the down part. These  larger differences for  the HOMO are
due to two  reasons: (i) the stronger mixing of  states in a structure
with lower  symmetry for Mn off-center, and (ii) the  small value for
the undoped  charge density  out of the  dot center. The  latter means
that when Mn  is off-center, the states  can suffer naturally stronger
perturbations.   Such  is  the  case  for the  HOMO-LUMO  states  with
off-center Mn. They  show larger voids or lower  values of the density
around Mn.

\subsection{ Analysis of Valence-Band Edge States and Occupied Mn Levels}

Exchange splitting of the conduction edge states has been discussed at
some  length in the  previous sections.   Here, we  look again  at the
splitting of the valence-band border  by looking at the nearest border
states.  This analysis  is useful to clarify the  breaking of symmetry
in CdTe  levels and Mn  \textit{d} levels due  to the Mn displacement  from the
central position.

\begin{figure}[t!]
\centering
\includegraphics[scale=0.82]{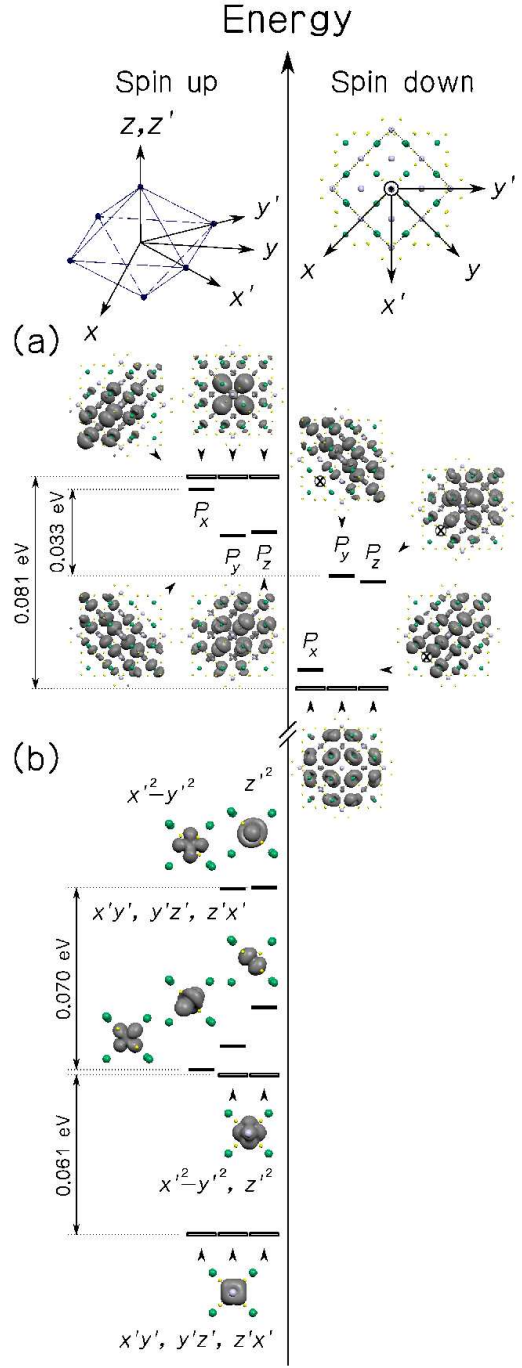}
\caption{\label{fig:VSI}(Color online) (a) Valence states close to the
Fermi energy for Mn in center  with empty lines; and for Mn off-center
in position II with full lines. The global densities are given nearest
to the corresponding states. The notation concerning atom labeling and
density  cuts   follows  the  one  given  in   previous  figures.  The
geometrical insets  show the  reference systems.  The  degeneracy gets
lower  for the  off-center  position, however  the quantum  dot states
$P_y$ and $P_z$ are nearly  degenerate.  (b) Mn $d$ states for center and
off-center position. We see the splitting between the triplet $t_{2}$
and   doublet  $e$  states   [see  next   figure  for   their  order
explanation]. }
\end{figure}

The valence edge  states for Mn in position I  are shown with empty
dashes in  Fig. \ref{fig:VSI} (a) for  up and down  spins.  The states
are three-fold degenerate  and we have commented at  some length about
their characteristics in the previous  section.  In this panel we show
also  the valence  edge  states for  Mn  in position  II  along the  $x$
direction of displacement.  The off-center Mn breaks the degeneracy of
these three states.  We  have identified the global characteristics of
these states,  although sometimes they have a  strong mixed character.
The  $P_y$ and  $P_z$  orbitals,  perpendicular to  Mn  shift, have  a
smaller splitting.  The global  $P_x$ suffers the largest splitting as
it  is parallel to  the Mn  displacement direction.   These electronic
levels for Mn in position  II explain naturally why the spin splitting
of up-down valence band edges is  reduced to half value, respect to Mn
at the center.

\begin{figure}[t!]
\centering
\includegraphics[scale=0.7]{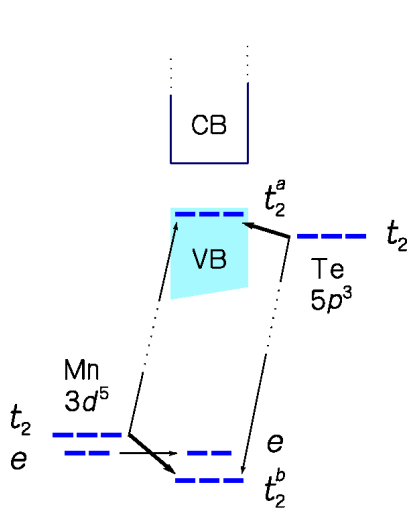}
\caption{\label{fig:scheme}(Color  online)  Branching  diagram  for  a
substitutional Mn  atom in CdTe  dot. The states of  substitutional Mn
atom (center) are originated in  the hybridization of vacant Cd states
$Va_{Cd}-t_{2}$ (right) with Mn-$d$ states (left).  }
\end{figure}

These  splittings are  correlated to  those of  Mn levels,  plotted in
Fig.  \ref{fig:VSI} (b).  For central  Mn  the occupied  \textit{d} levels  are
grouped in two  sets: the first has $d_{x'^2-y'^2}$  and $d_{z'^2}$, the
second  gets the  cross \textit{d}  levels $x'y'$,  $y'z'$, and  $z'x'$.  These cross
orbitals  have lower  energy  which is  typical  for  a Cd  vacancy
$Va_{Cd}$ interacting  with the tetrahedrally split $Mn-d$ electrons [see
branching scheme in  Fig. \ref{fig:scheme}]. The removal of  a Cd atom
creates  a  vacancy  $Va_{Cd}$.  The  $t_2$  levels  of  the  $Va_{Cd}$
hybridize with the $t_{2}$ states  of Mn atom and produce the bonding
$t_{2}^b$  and antibonding $t_{2}^a$  states. As  the Mn  states are
very  low   in  energy  the  $t_{2}^b$  states   resembles  much  the
$Mn-t_{2}$  states and  cross  over the  $Mn-e$  states. While  the
$t_{2}^a$ states remain close to the QD semiconductor states. For Mn in position II, the $d$ levels shift upwards in energy, typical for the smaller CdTe density around Mn, as seen in the section about site dependence of exchange constants. The $e$ levels remain degenerate. The $t_2^b$ states ($x'y'$, $y'z'$, and $z'x'$) split although they still are groupable together. The orbital $z'x'$ orthogonal to the Mn displacement has the highest energy. The orbitals $x'y'$ and $y'z'$ remain at lower energy, the lowest energy for the $x'y'$ orbital with the lobes in the Mn displacement direction from the center. 

We  want to  stress finally  that the  Mn has  a much  larger exchange
splitting of several eV that polarizes the CdTe  levels in much smaller amount, in the order of several hundredths of eV.  This
spin polarization of CdTe in  NCs depends on the Mn position. Although
the splitting of CdTe levels  that interact strongly with Mn is almost
independent of  position ($P_x$), the  other levels ($P_y$  and $P_z$)
remain  almost  unaffected.  This  finding  means  that  the  exchange
splitting of the up-down CdTe  valence edge amounts to nearly less than
half of the value with central Mn.

\section{Nanocrystals with two Mn impurities}
\label{nanotwo}

We  investigate  (Cd,Mn)Te NCs  doped  with  two  Mn impurities.   The
substitutional energy differences $\Delta  E_2$ are calculated for the
following reaction:

\begin{equation}
\begin{array}{ccc}
\textrm{CdTe}\;\textrm{NC}+2\textrm{Mn}^{+2} & \longrightarrow & \textrm{CdTe:Mn}\;\textrm{NC}+2\textrm{Cd}^{+2}.
\end{array}
\end{equation}

\begin{figure}[t!]
\centering
\includegraphics[scale=0.8]{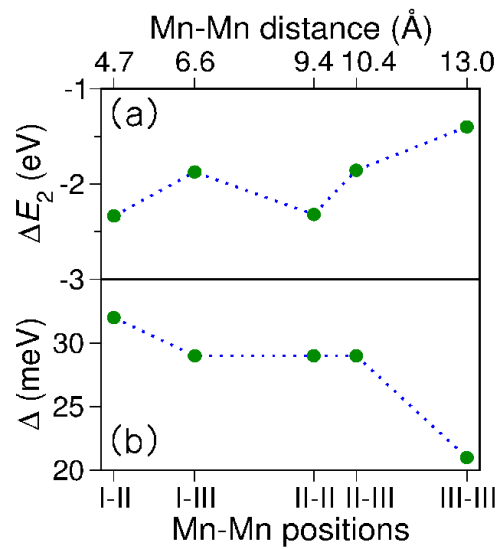}
\caption{\label{fig:E02}(Color online) (a) Substitutional energy of two Mn positions.
(b) Interaction energy $\Delta$ for  various Mn-Mn pairs. It shows the
difference  between the  previous total  Mn-Mn energy  and the  sum of
single Mn energies in the NC.}
\end{figure}

The calculated  differences against the Mn positions  and the distance
between Mn  atoms are displayed  in Fig. \ref{fig:E02}.  When  both Mn
impurities occupy sites I-II and  II-II, the energies $\Delta E_2$ are
larger than for positions I-III  and II-III.  This difference is about
0.5 eV.  The  substitutional energy of Cd decreases  with the presence
of Mn  atoms at  surface sites III.   This trend follows  the previous
energy    differences   $\Delta    E_1$   for    a   single    Mn   in
Fig.  \ref{fig:esfera}, where position  III is  the most  stable.  The
interaction energy is the difference between the previous $\Delta E_2$
Mn-Mn energy and the sum of single Mn energies $\Delta E_1$ in the NC.
We  plot also  the interaction  energy ($\Delta$)  between the  two Mn
impurities in  Fig. \ref{fig:E02} (b).  The changes  of the interaction
energy $\Delta  $ due to Mn  positions are an order  of magnitude lower
than those of the substitutional energy  $\Delta E_2$.  The interaction $\Delta
$ is 32 meV  for Mn-Mn atoms in sites I-II and  21 meV in III-III.  It
seems that the  closer the Mn impurities the  larger their interaction
energy.   However,  the dependence  of  $\Delta  $  with the  distance
between the two Mn impurities is  almost flat except when Mn atoms are
next-nearest neighbors or close to the surface.

\begin{figure}[t!]
\centering
\includegraphics[scale=0.75]{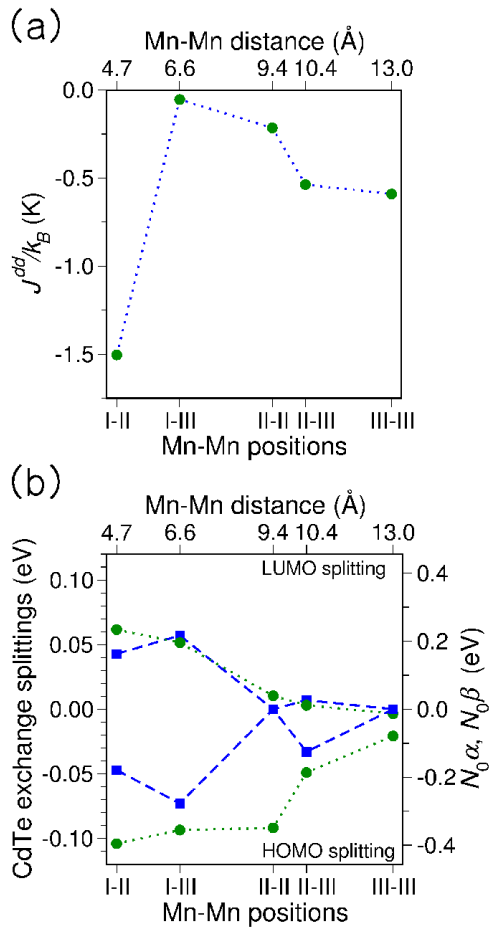}
\caption{\label{fig:ec-2}(Color online) The (a) \textit{d-d} exchange constant together with the (b) \textit{sp-d} exchange constants ($N_0\alpha$ and $N_0\beta$) for several Mn-Mn positions. The $J^{dd}$ are given in K to compare with the bibliography. Exchange splittings follow the notation of Fig. 4: Circle marks with dotted lines (green) denote FM alignments; and square marks with dashed lines (blue), AFM alignments.}
\end{figure}

To  a  lesser  extent  these  energies $\Delta E_2$ and $\Delta$ also  depend  on  the  magnetic
configuration between  Mn magnetic  moments. We consider  two magnetic
coupling cases  for Mn  moments: ferromagnetic in  which $\mu_{QD}=10$
$\mu_{B}$,  and  antiferromagnetic  in which  $\mu_{QD}=0$  $\mu_{B}$.
Independently of the Mn positions within the QD, the Mn magnetic moments
in the lowest-energy state are coupled antiferromagnetically.  This is
shown in Fig.~\ref{fig:ec-2}  using the \textit{d-d} exchange constant
for   various   Mn-Mn  positions   instead   of   the  FM-AFM   energy
differences. In fact, we  calculate the \textit{d-d} exchange constant
from the FM-AFM energy  differences for various Mn-Mn positions. These
constants  $J^{dd}$ are  depicted  in Fig.~\ref{fig:ec-2} (a).   The
largest value in modulus is roughly 1.5  K for Mn atoms on sites I and
II. It  is about  four times  smaller than the  absolute value  of the
experimental         exchange         constant         in         bulk
$\rm{Cd}_{0.95}\rm{Mn}_{0.05}Te$, 6.1  K.  In addition we  can see the
dependence with  the Mn-Mn  sites.  For Mn-Mn  intermediate distances,
the exchange  constants decrease even  further and they  are following
the AFM-FM  splitting.  Note that for  the Mn atoms in  the center and
surface positions this splitting seems to be very small.

We  have  also  examined  the  $d$-levels for  several  of  these  Mn-Mn
configurations, although they are not  given here. Due to symmetry in
QDs, the  $d$-levels in position  I and II  are close, but  in different
energy  panels [as  seen  in previous  Fig.~\ref{fig:VSI}]. This  means that  the
symmetry considerations are more relevant in QDs to understand
the  Mn-Mn   interaction  that   higher-order  mechanisms,   such  as
superexchange  or  double  exchange.   We  must  remember  that  such
mechanisms are  important for DMS bulks  where the $d$-levels  are at the
same energies,  due the  equal density around  Mn atoms.  These symmetry
considerations and the smaller dot density around Mn atoms explain the
lower coupling $J^{dd}$ constants.

The $N_0\alpha$  and $N_0\beta$ exchange  constants are shown  in Fig.
\ref{fig:ec-2}  (b).    They  are  calculated   for the  ferromagnetic
(circles)  and antiferromagnetic  (squares) states  for  several Mn-Mn
positions;  the dopant  concentration  is $x=2/19\sim0.1$.  As in  the
previous  section,  the exchange  constants  get  closer  to the  bulk
experimental  values when  one of  the Mn  impurities occupies  the NC
center  (I-II and  I-III), and  they  tend to  zero when  the Mn  ions
separate  from  each other  and  approach  the  surface. The  exchange
constants for antiferromagnetic cases with high symmetry are also zero
because the down  and the up electrons are fully  degenerated.  When comparing
Fig.  \ref{fig:dos2} with  Fig.  \ref{fig:ec-2}, we  conclude that  the
largest  exchange values are  obtained for  NCs doped  only with  a Mn
impurity in the central site.  The relative orientation of $N_0\alpha$
and $N_0\beta$ is the same as before. However their values decrease by
at least 0.2 eV. This difference can be ascribed to the second Mn atom
that  cancels partially  the coupling  to the  conduction  and valence
bands in the neighborhood of the first Mn.

\subsection{Valence Edge States}

\begin{figure}[h!]
\centering
\includegraphics[scale=0.8]{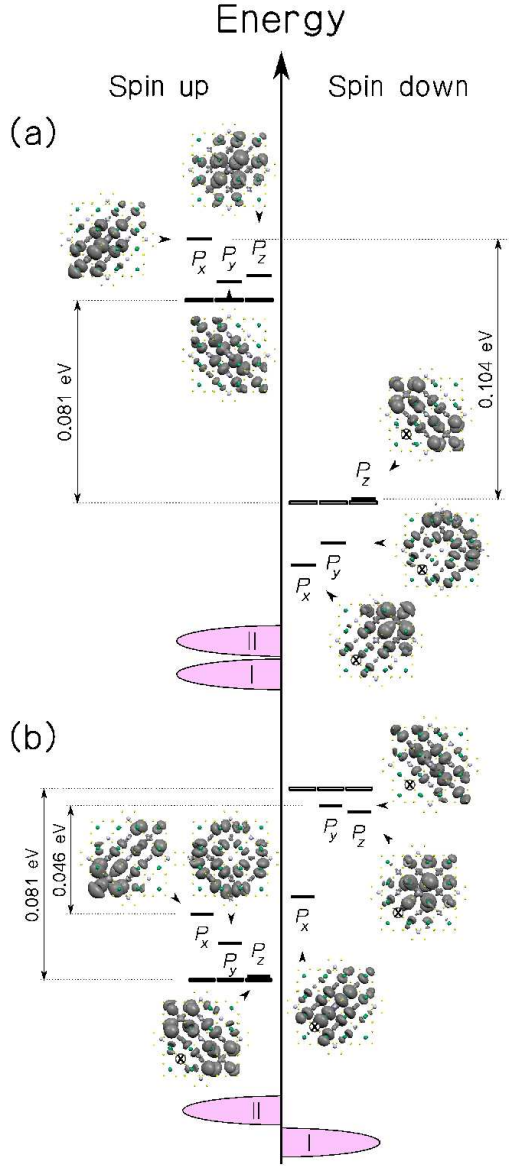}
\caption{\label{fig:VSI-II}(Color  online) Valence  states  closest to
the Fermi energy for Mn in  configuration I-II with full lines for (a)
FM and  for (b) AFM  couplings. The empty  dashes refer to  a previous
case with  Mn in  a central position.  The global densities  are given
nearest  the  corresponding   states.  The  notation  concerning  atom
labeling  and   density  cuts  follows  the  one   given  in  previous
figures. At the bottom, a scheme  of Mn magnetic moments is shown.  For
the ground AFM case the valence states split according to center Mn in
position I, however  the degeneracy follows closely the  off-center Mn
in position II.}
\end{figure}

\begin{figure}[h!]
\centering
\includegraphics[scale=0.8]{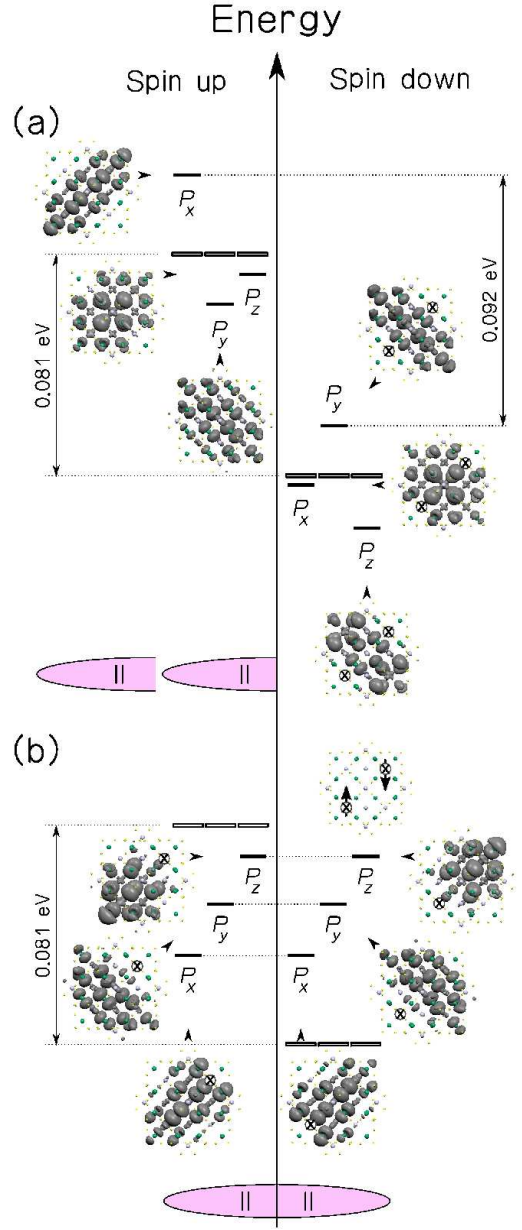}
\caption{\label{fig:VSII-II}(Color  online) Valence states  closest to
the Fermi  energy for  Mn in symmetric  configuration II-II  with full
dashes for (a) FM and for (b) AFM couplings.  They follow the notation
of the previous  figure.  For off-center positions  the valence states
break the degeneracy even further.}
\end{figure}

The valence  states around the edges  allow us,  like in the
case of a single Mn impurity analysis, to understand the origin of the
exchange in  the CdTe  states. As  it has been  done before,  they are
plotted  for   the  asymmetric   Mn-Mn  configuration  I-II   in  Fig.
\ref{fig:VSI-II}    and    for   the    symmetric    one   II-II    in
Fig. \ref{fig:VSII-II}.  However, due  to magnetic coupling  there are
more options than in one Mn  impurity case. Thus the panels (a) denote
the FM Mn alignments; and the panels (b), the AFMs.

We shall  see that  for the results  of both figures,  a perturbation
theory of the central Mn  case rationalizes also the level breaking as
due  to symmetry and  magnetic configuration.  We therefore  start the
discussion by  considering the AFM case I-II  in Fig. \ref{fig:VSI-II}
(b).

We see that  the exchange splitting of the CdTe aligns  with the Mn in
position I.   Let us consider  that a Mn  in position II  perturbs the
levels to  first order, as  in previous one  Mn impurity. Now,  we can
focus  on the  change in  splitting of the  three-fold  degenerate levels
using the  previous results. In  this way, the down  CdTe valence-band
edge levels are  $P_x$ at low energy, and  nearly degenerate $P_y$ and
$P_z$  at  higher   energies.   For  the  up  levels   this  order  is
reversed.  This  order  is  also  followed by  our  computations.  The
expected  and calculated levels  are in  excellent agreement  and thus
confirm that a perturbative approach  applies to the splitting of CdTe
levels by central Mn.

However,  the  case  AFM I-II we  have  discussed, in  which  the  perturbative
approach can be applied, is not the only case of
physical  interest. In  the  FM case,  Fig.  \ref{fig:VSI-II} (a)  for
example, both  Mn reinforce  their splitting of  the CdTe  states. The
value is close to the sum of both I and II splittings for
the Mn  impurities on their own.  We shall not go  into further detail
here.

For the  case II-II we only draw  attention to the fact  that there is
strong  interaction  for  one  up  Mn  impurity  with  the  CdTe  down
states. We  can see the empty  space of up (down)  orbitals around the
down  (up)  Mn  in  Fig.   \ref{fig:VSII-II}  (b)  as  it  expels  the
surrounding charge.  We want to  stress here that  for the FM  case in
Fig. \ref{fig:VSII-II}  (a) we need  to go beyond the  previous simple
perturbative analysis.  In spite of the strong  perturbation, it means
that  it is  more difficult  to interpret  the data  as coming  from a
perturbation due to a second impurity. It remains true that there is
a clear increase of the splitting of $P_x$ orbital aligned along the
corresponding Mn-Mn direction.

\section{Conclusions}
\label{conclusion}

In summary,  we have investigated the electronic and  magnetic properties of
(Cd,Mn)Te nanocrystals of spherical  shape with the density functional
theory. The embedded Mn impurities substitute Cd atoms in the zinc
blende  lattice. The QDs  are $\sim2$  nm in  diameter, centered  on a
cation site (Cd),  and they have a  total of 107 atoms. The  Cd and Te
dangling bonds  on the surface are passivated  by pseudohydrogen atoms
($\rm{H}^\ast$),  which are  the  simplest model  for  a real  organic
ligand.   We  have studied  two  doping   situations:  NCs
including one and two Mn impurities.

In the first case we find that the configuration with the Mn atom near
the surface  presents the  lowest energy. The  Mn impurity  introduces five
\textit{d}-type  spin-up  electrons  within  the NC,  thus  the  total
magnetic moment associated with the QD is 5 $\mu_B$, as expected.  The
local Mn magnetic moment is 4.65 $\mu_B$. It is smaller than 5 $\mu_B$
because    of    the    \textit{s,p-d}   hybridization.    Also    the
Mn-nearest-neighbor   Te   sites    show   small   magnetic   moments
antiferromagnetically coupled  to the Mn moment.  Furthermore, we look
at  the hybridization  in the  density  of states.   We calculate  the
\textit{sp-d}  exchange constants for  various Mn  locations. The exchange 
constants are comparable to the bulk ones when the impurity is at the center, 
and they  go to zero when the Mn  is close to the NC surface. 
Then, for central Mn we introduce spatially the HOMO and LUMO
states to verify  that the largest change is for  the down HOMO, which
also justifies the antiferromagnetic alignment of the nearest-neighbor
Te atoms.   Additionally, we check the  role of symmetry  by looking at
the CdTe  levels near the  valence-band edge.  This allows us  a clear
interpretation of the origin of exchange splitting. The $P_x$ levels in the
Mn diplacement direction suffer the stronger hybridizaton, while the $P_y$ and $P_z$ levels remain almost unsplit. Thus, the exchange contants are roughly
half of the central Mn case.

In the  second case  we calculate the  total ground-state energy  as a
function of  Mn-Mn positions and  the magnetic configuration  of their
local magnetic moments. In the minimum-energy state the Mn dopants are
on the  surface, as far  apart as possible,  and antiferromagnetically
coupled.   We calculate  the \textit{sp-d}  and  \textit{d-d} exchange
constants  for various  Mn-Mn  locations. We  find  that the  exchange
constants   $N_0\alpha$   and  $N_0\beta$   are   comparable  to   the
corresponding  bulk values  when impurities  occupy  central bulk-like
positions (I-II). They  tend to zero as they  separate from each other
or approach the surface. As  for the exchange $\vert J^{dd}\vert$, its
largest value  is for  the Mn ions  placed in central  bulk-like sites
(I-II), and is four times  smaller than in the bulk with similar
Mn  concentration.    A  similar   analysis  of  states   nearest  the
valence-band edge  demonstrates that  a perturbative approach
seems to be valid, with exception of the II-II FM configuration.

\acknowledgments{

 This  work  was  supported  by  the  Basque  Government  through  the
NANOMATERIALS project (IE05-151)  under the ETORTEK Program (NANOMAT),
Spanish Ministerio  de Ciencia y  Tecnolog\'ia (MCyT) of  Spain (Grant
No.  Fis  2004-06490-CO3-00  and  MONACEM project)  and  the  European
Network of Excellence  NANOQUANTA (NM4-CT-2004-500198).  The computing
resources from  the Donostia  International Physics Center  (DIPC) are
gratefully acknowledged.  Prof. A. K. Bhattacharjee is acknowledged for
comments  on the manuscript.   Carlos E.-A.  wants to  thank N.  Gonz\'alez for
support on the plotting routines during this work.  }

\newpage




\clearpage
\end{document}